\documentclass[aps,prd,twocolumn,groupedaddress,asmsymb,amsmath,amssymb,longbibliography]{revtex4-1}

\usepackage{graphicx}
\usepackage{calligra}
\usepackage{upgreek}
\usepackage{color}
\usepackage{inputenc,caption}
\usepackage{stackrel}
\usepackage[usenames,dvipsnames]{xcolor}
\usepackage{subcaption}

\numberwithin{equation}{section}

\DeclareMathAlphabet{\mathpzc}{OT1}{pzc}{m}{it}

\newcommand{\be}{\begin{equation}}
\newcommand{\ee}{\end{equation}}
\newcommand{\bea}{\begin{eqnarray}}
\newcommand{\eea}{\end{eqnarray}}
\newcommand{\lb}{\label}

\newcommand{\bu}{{\bf u}}

\newcommand{\bx}{{\bf x}}

\newcommand{\bI}{{\bf I}}

\newcommand{\bK}{{\bf K}}

\newcommand{\bS}{{\bf S}}

\newcommand{\bV}{{\bf V}}

\newcommand{\bomega}{{\mbox{\boldmath $\omega$}}}

\newcommand{\btau}{{\mbox{\boldmath $\tau$}}}

\newcommand{\grad}{{\mbox{\boldmath $\nabla$}}}
\newcommand{\bdot}{{\mbox{\boldmath $\cdot$}}}
\newcommand{\bdots}{{\mbox{\boldmath $:$}}}
\newcommand{\bzed}{{\mbox{\boldmath $0$}}}
\newcommand{\btimes}{{\mbox{\boldmath $\times$}}}
\newcommand{\bn}{\hat{{\mbox{\boldmath $n$}}}}

\definecolor{forestgreen}{rgb}{0.13, 0.55, 0.13}
\definecolor{bondiblue}{rgb}{0.0, 0.58, 0.71}
\definecolor{paleblue}{rgb}{0.4, 0.7, 1}
\definecolor{palered}{rgb}{1, 0.48, 0.25}

\newcommand{\black}[1]{\textcolor{black}{#1}}

\begin{document}


\title{Weak-Strong Uniqueness and Extreme Wall Events at High Reynolds Number}


\author{Gregory Eyink$^{1,2}$}
\email[]{eyink@jhu.edu}
\author{Hao Quan$^1$}
\affiliation{$^1$Department of Applied Mathematics \& Statistics, The Johns Hopkins University, Baltimore, MD, 21218, USA}
\affiliation{$^2$Department of Physics \& Astronomy The Johns Hopkins University, Baltimore, MD 21218, USA}



\begin{abstract}
Singular or weak solutions of the \black{incompressible} Euler equations have been hypothesized to account for anomalous 
dissipation at very high Reynolds numbers and, in particular, to explain the d'Alembert 
paradox of non-vanishing drag. A possible objection to this explanation is the mathematical 
property called ``weak-strong uniqueness'', which requires that any admissable weak solution 
of the Euler equations must coincide with the smooth Euler solution for the same initial data.
As an application of the Josephson-Anderson relation, we sketch a proof of conditional 
weak-strong uniqueness for the potential Euler solution of d'Alembert within the class of strong inviscid limits. 
We suggest that the mild conditions required for weak-strong uniqueness are, in fact, physically violated by violent 
eruption of very thin boundary layers. We discuss observational signatures of these extreme events and 
explain how the small length-scales involved could threaten the validity of a hydrodynamic description. 
\end{abstract}

\pacs{?????}

\maketitle


\section{Introduction}\lb{sec:intro}

Various empirical observations suggest that drag and dissipation are non-vanishing 
for turbulent flows of incompressible fluids at low Mach number, even in the limit of infinite Reynolds number. 
Evidence provided by laboratory experiments necessarily involves flow interactions 
with solid surfaces, such as asymptotic non-zero values of drag coefficients for flow past 
bodies: \cite{frisch1995turbulence}, \S 5.2.
It was conjectured by Onsager \cite{onsager1949statistical,eyink2024onsager}
that such turbulent ``anomalous dissipation'' can be explained by singular or weak solutions of ideal 
Euler equations and recent works have developed this theory for flows interacting 
with solid walls. In particular, a solution of the famous d'Alembert paradox 
\cite{dalembert1749theoria,dalembert1768paradoxe}
can be formulated 
within this approach, according to which the drag on the body in the infinite Reynolds number 
limit is produced by some weak Euler solution with vorticity in the wake, supplanting the potential
flow of d'Alembert that exerts no drag \cite{hoffman2010resolution,quan2024onsager}. 

A challenge for this proposed solution is the notion of ``weak-strong uniqueness''
in the mathematical theory of partial differential equations \cite{wiedemann2018weak}. 
Since this concept seems to be not widely known to fluid physicists, we shall summarize it here 
succinctly and explain the difficulty it poses for Onsager's theory. We shall also sketch a simple proof for the case of the d'Alembert flow 
\cite{quan2025weak}, which exploits the Josephson-Anderson relation recently derived 
for such external flows \cite{eyink2021josephson,eyink2021jaerratum}. The crucial point of our work 
is that weak-strong uniqueness holds only conditionally for fluids interacting 
with solid boundaries. If these sufficient conditions do not hold, then the challenge to 
Onsager's theory from weak-strong uniqueness is eliminated. The sufficient conditions 
appear rather mild, so that, conversely, any 
violation of weak-strong uniqueness requires quite extreme events. In fact, we shall argue 
that these events are so singular that they could threaten the very validity of a continuum hydrodynamic description.
It is the main purpose of this paper to delineate the striking signatures of this breakdown
in order to facilitate further investigations by laboratory experiment and by numerical simulation.  
Our discussion may be useful also for mathematicians as a short summary 
of relevant empirical observations. 

\section{What is Weak-Strong Uniqueness?}\lb{sec:wsuniq} 

An informal statement of weak-strong uniqueness is as follows: {\it If a smooth Euler solution 
${\bf u}$ exists on spacetime domain $\Omega\times [0,T]$ with initial data 
${\bf u}(\cdot,0)={\bf u}_0$, then any generalized Euler solution on the same domain 
with the same initial data  must coincide with that smooth solution, so long as the 
generalized solution satisfies some modest “admissibility” requirement.} 
A common example of an admissibility requirement is the global energy bound
\be {{1}\over{2}}\int_\Omega |{\bf u}({\bf x},t)|^2 \, dV\leq {{1}\over{2}}\int_\Omega |{\bf u}_0({\bf x})|^2 \, dV, \qquad \forall t\geq 0, \lb{admissable} \ee
which is an expected consequence of energy dissipation. 
In the spatial domains $\Omega={\mathbb R}^d$ or ${\mathbb T}^d$ {\it without boundaries} 
for $d\geq 2$, any locally dissipative weak Euler solution (in particular any weak 
solution obtained as the strong inviscid limit of Navier-Stokes 
solutions) enjoys the weak-strong uniqueness property. Such strong limits exist 
under certain reasonable conditions \cite{drivas2019remarks}  but even more generalized 
notions of ``Euler solution'' are guaranteed to exist in the inviscid limit (at least along a subsequence of viscosities) 
and also enjoy weak-strong uniqueness. These include dissipative Euler solutions ``in the sense of Lions''  \cite{lions1996mathematical,bardos2013mathematics}
and measure-valued Euler solutions \cite{brenier2011weak}. See \cite{wiedemann2018weak}
for a very lucid mathematical review of these and related results. A common view 
expressed in the mathematics literature is 
that weak-strong uniqueness is a necessary requirement for any ``reasonable'' notion of generalized 
Euler solution. The underlying presumption is that a smooth Euler solution, whenever it exists, 
must be the ``physical'' solution in the inviscid limit.   

One important implication of \black{weak-strong uniqueness} is on existence of blow-up or finite-time singularities
for smooth Euler solutions. Indeed, weak-strong uniqueness for the inviscid limit solution 
implies that viscous energy dissipation must necessarily vanish over any finite time interval 
for which the smooth Euler solution exists \cite{eyink2008dissipative,bardos2013mathematics}. 
Conversely, appearance of anomalous energy dissipation over a finite time interval for a 
flow with smooth initial data requires that the  \black{incompressible} Euler solution with that initial data cannot remain 
smooth over the interval. A typical example of such initial data is the Taylor-Green vortex 
in periodic domain ${\mathbb T}^3$ \cite{taylor1937mechanism},
although existing evidence for anomalous dissipation in this flow \cite{fehn2022numerical} seems 
less compelling to us than corresponding evidence for wall-bounded flows. In any case, a 
finite-time Euler singularity can be rigorously ruled out for certain smooth initial data, e.g. 
for any smooth data in a 2-dimensional spatial domain and for the 
stationary potential data in the 3-dimensional d'Alembert flow. Thus, finite-time blow-up
of smooth Euler solutions cannot be the general route to turbulence and anomalous dissipation
\black{in incompressible fluid flows}. 

\section{The New D'Alembert Paradox}\lb{sec:alembert} 

Perhaps the most famous example of a smooth Euler solution which exists globally in time 
in a three-dimensional space domain is the solution of d'Alembert for potential flow past 
a smooth body. The context is a rigid body occupying a closed set 
$B\subseteq{\mathbb R}^3$ with a smooth boundary $\partial B$ immersed 
in a fluid filling an infinite-volume domain $\Omega={\mathbb R}^3\backslash B$
with velocity ${\bf V}(t)$ at infinity. Alternatively, by Galilean invariance, one
may consider a rigid body moving with translational velocity $-{\bf V}(t)$ 
through a fluid at rest at infinity, but we find it more convenient to 
discuss the problem in the rest frame of the body.  There is a unique 
potential flow solution ${\bf u}_\phi={\boldsymbol\nabla}\phi$ of the incompressible Euler equation
$$ \partial_t{\bf u}_\phi+{\boldsymbol\nabla}{\boldsymbol\cdot}({\bf u}_\phi{\bf u}_\phi+p_\phi{\bf I})={\bf 0}, \quad {\boldsymbol\nabla}{\boldsymbol\cdot}{\bf u}_\phi=0, $$
where the the potential $\phi(\cdot,t)$ is defined at each time instant 
$t\in [0,T]$ as 
the solution of the Neumann boundary value problem
\begin{eqnarray}
\triangle\phi(\bx,t)=0, &&\,  \bx \in \Omega,    \cr
\frac{\partial\phi}{\partial n}(\bx,t)=0, &&\,  \bx\in \partial B, \cr
\phi(\bx,t)\sim {\bf V}(t)\bdot\bx, &&\,  |\bx|\to\infty, 
\end{eqnarray}
unique up to a spatial constant, and with pressure $p_\phi$ likewise 
determined up to a spatial constant from the Bernoulli equation
$$ \partial_t\phi+ \frac{1}{2}|{\bf u}_\phi|^2 + p_\phi = 0. $$
If the velocity ${\bf V}(t)$ is smooth as function of time, then this 
unique potential Euler solution ${\bf u}_\phi$ is smooth on the space-time 
domain $\Omega\times [0,T].$

The force exerted on the body $B$ by the potential Euler flow past it is given 
instantaneously by the integral 
$${\bf F}_\phi(t) = -\int_{\partial B} p_\phi(\cdot,t) \bn \, dA, $$
where $\bn$ is the unit normal on the body surface pointing into the fluid. 
For simplicity, we have taken fluid density $\rho\equiv 1$ here and throughout.  
The famous result of d'Alembert \cite{dalembert1749theoria,dalembert1768paradoxe}
for the stationary potential
flow with constant fluid velocity $\bV(t)\equiv {\bf V}$ is that 
the force on the body vanishes identically, ${\bf F}_\phi\equiv\bzed.$ 
This result can be generalized to time-dependent potential flow as long as $\bV(t)$
is bounded in time and, thus, the fluid impulse \cite{lighthill1986informal} 
$$ \bI_\phi(t) = -\int_{\partial B} \phi(\cdot,t) \bn \, dA $$
is also bounded. Since ${\bf F}_\phi(t) = -d\bI_\phi(t)/dt,$
it then follows easily that long-time average force must vanish:
$$\langle {\bf F}_\phi\rangle=
\lim_{T\to\infty} \frac{1}{T} \int_0^T {\bf F}_\phi(t)\,dt ={\bf 0}. $$
In addition, the power dissipated by fluid drag force, ${\mathcal W}_\phi(t)={\bf F}_\phi(t)\bdot\bV(t),$
can be written likewise as a total time-derivative by noting that $\bI_\phi(t)={\mathbb M}_A\bdot\bV(t)$
in terms of the ``added mass tensor'' ${\mathbb M}_A$ of the rigid body
\cite{lighthill1986informal}. In that case, 
${\mathcal W}_\phi(t)=-(d/dt)\left(\frac{1}{2}{\mathbb M}_A\bdots\bV(t)\bV(t)\right),$ and 
one similarly obtains  
$$\langle {\mathcal W}_\phi\rangle=
\lim_{T\to\infty} \frac{1}{T} \int_0^T {\mathcal W}_\phi(t)\,dt =0. $$
This vanishing seems to contradict common experience that 
drag forces are non-zero and it led to the famous ``d'Alembert paradox.''
The accepted resolution \cite{stewartson1981d} of this original paradox is that 
proposed by Saint-Venant and Navier, which is that molecular fluids have a viscosity 
$\nu>0$ and are governed by the Navier-Stokes equation 
$$ \partial_t{\bf u}+{\boldsymbol\nabla}{\boldsymbol\cdot}({\bf u}{\bf u}+p{\bf I}
-\nu\grad\bu)=\bzed, \quad {\boldsymbol\nabla}{\boldsymbol\cdot}{\bf u}=0, $$
which predicts a non-vanishing time-average drag dissipation 
$\langle{\mathcal W}^{Re}\rangle\neq 0$ for any finite value of the Reynolds
number $Re=VD/\nu.$ Here $D$ is diameter of the body $B.$

However, a {\it new d'Alembert paradox} arises if one considers the limit $Re\to\infty$
\black{\cite{stewartson1981d}}.
There is substantial empirical evidence that drag remains non-vanishing in this limit,
for example, the apparent non-zero asymptotic values of drag coefficients for solid bodies of various shapes. Nevertheless, any such limits must be described by generalized Euler 
solutions \cite{lions1996mathematical,diperna1987oscillations} or, under reasonable 
assumptions, by weak Euler solutions as conjectured by Onsager \cite{drivas2019remarks} 
and \black{a suitable generalization of} the admissability condition \eqref{admissable} furthermore must hold. 
Weak-strong uniqueness would require that the generalized Euler solution obtained 
in the limit must coincide with the smooth Euler solution of d'Alembert, 
suggesting that time-average drag must vanish in the limit $Re\to\infty.$
To argue for this conclusion, one may consider the situation where the fluid starts at rest with ${\bf V}(0)={\bf 0}$ and then smoothly accelerates to a velocity ${\bf V}(t)$ bounded in time.  Taking $T$ so large that 
$\left|(1/T)\int_0^T {\mathcal W}_\phi(t)\,dt\right|<\epsilon$, then weak-strong uniqueness implies \cite{quan2025weak} that 
\be \lim_{Re\to\infty}\left|(1/T)\int_0^T {\mathcal W}^{Re}(t)\,dt\right|<\epsilon. \lb{dragzero} \ee 
Note, however, that tabulated drag coefficients are not so well-characterized experimentally, e.g. 
the body may have accelerated through fluid with small turbulent fluctuations and not initially at rest. 
Thus, it is not completely clear that weak-strong uniqueness contradicts the observations of non-vanishing drag. 

Beyond simple drag coefficients, however, more carefully controlled laboratory and numerical experiments have 
been performed to measure force histories and other characteristics for the case of accelerated bodies. 
This is particularly true for the classical problem of 
the impulsively accelerated cylinder initiated in the 1925 work of Prandtl \cite{prandtl1925magnuseffekt} 
and since extensively studied experimentally \cite{prandtl1931hydro,schwabe1935druckermittlung,honji1969unsteady,taneda1977visual,bouard1980early,loc1985numerical,nagata1985unsteadyA,nagata1985unsteadyB}
and computationally 
\cite{payne1958calculations,collins1973flow,chang1991vortex,koumoutsakos1995high,mittal2008versatile,chatzimanolakis2022vortex}. 
This case is interesting because it corresponds in the body frame to the Navier-Stokes solution ${\bf u}^{Re}$ with initial data given by d'Alembert's stationary potential flow, ${\bf u}^{Re}(\cdot,0)
={\bf u}_\phi$. Notice for this initial data that ${\boldsymbol \Gamma}=\bn\btimes\bu_\phi\neq \bzed$ 
corresponds to a singular vortex sheet at the body surface. A delicate aspect of such impulsively accelerated flows is that the very earliest stages cannot be described hydrodynamically. 
Kinetic theory shows that slightly modified initial conditions and small slip at the boundary are necessary to match the solutions of Boltzmann equation and Navier-Stokes equation.   
These complications make the comparison of the Prandtl problem with hydrodynamic theory a bit more 
involved, as we discuss carefully in Section \ref{sec:proof}. It is easier to analyze theoretically 
the case of gradually accelerated bodies, a case which has been studied also quite extensively. 
We mention here just a few papers on experiment 
\cite{taneda1971unsteady,pullin1980some,grift2019drag,reijtenbagh2023drag} 
and simulation \cite{koumoutsakos1996simulations,homann2013effect,xu2015start,sader2024starting}, 
out of a huge literature. None of these various studies show any clear evidence of the high-Reynolds 
number flow converging to d'Alembert's potential flow solution. Although further careful
studies at much higher Reynolds numbers are required to provide more convincing data, we 
shall explain next a physical mechanism that can violate weak-strong uniqueness and 
thus permit a singular Euler solution in the inviscid limit distinct from d'Alembert's smooth potential 
solution.

\section{Weak-Strong Uniqueness Can Fail With Solid Walls}\lb{sec:fail}  

The traditional view in the mathematics community has been that weak-strong uniqueness
is a necessary requirement for any ``reasonable" concept of generalized solution, but recent
mathematical works on flows with solid boundaries have called into question this presumption.
A sigificant step in this direction is the result on non-uniqueness of weak Euler 
solutions in an annular domain for piecewise-smooth initial data with an interior vortex sheet, 
established by convex integration methods \cite{bardos2014non}. The initial data in this 
example is a stationary strong Euler solution, which co-exists with infinitely many admissible 
weak solutions that exhibit spreading mixing layers evolving from the initial vortex sheet. This 
is not a perfect counterexample to weak-strong uniqueness, however, because the stationary Euler 
solution has a jump discontinuity in its velocity field. A much cleaner example has been
established recently, also by exploiting convex integration methods developed for vortex-sheet initial data \cite{szekelyhidi2011weak}. In the case of a plane-parallel channel with initial data 
given by constant plug flow,  ${\bf v}_0={\bf U}$, it has been proved that the obvious stationary solution
of Euler (${\bf v}(\cdot,t)\equiv {\bf U}$, plug flow) co-exists with infinitely-many admissible weak 
Euler solutions in which the surface vortex sheet separates and mixes in the interior \cite{vasseur2023boundary}. 
Since plug flow is $C^\infty$ (in fact, analytic), this is a clear mathematical counterexample 
showing how weak-strong uniqueness can be violated by separation at the boundary. Such weak Euler 
solutions with boundary separation may not be physically realizable in channel flow with smooth, 
plane-parallel walls, as we discuss in Section \ref{sec:channel} below, but we argue that weak-strong uniqueness 
can be violated by this mechanism in physical examples such as the d'Alembert potential flow. 

Although clear counterexamples were found only recently, it has been known for awhile that standard proofs 
of weak-strong uniqueness break down in flows with solid walls unless additional conditions on the 
solutions beyond \eqref{admissable} are imposed. For example, \cite{bardos2013mathematics} have shown that 
weak-* limits of Navier-Stokes solutions in a bounded domain $\Omega$ with stick boundary conditions 
are dissipative Euler solutions in the 
sense of Lions on $\bar{\Omega}\times [0,T) $ and thus satisfy weak-strong uniqueness, but only under 
the following condition: the wall shear stress or skin friction ${\boldsymbol \tau}_w^\nu:=2\nu \bS^\nu\bdot\bn$ 
on the boundary must satisfy 
\be {\mathcal D}-\lim_{\nu\to 0} {\boldsymbol \tau}_w^\nu= {\bf 0},  \lb{BTcond} \ee 
with convergence in the sense of distributions on the domain $\partial\Omega\times [0,T].$ 
The same result was established in \cite{bardos2013mathematics} for Navier-slip boundary 
conditions as well, as long as the normalized slip length $\lambda/D\to 0$ as $Re\to \infty.$
It was also proved 
in \cite{bardos2013mathematics} that \eqref{BTcond} is implied by the condition of 
vanishing dissipation in a neighborhood of the boundary 
\be \lim_{\nu\to 0}\int_0^T \int_{\Omega_{c\nu}} \nu|{\boldsymbol\nabla}{\bf u}^\nu|^2\, dV\, dt=0, 
\lb{Kcond} \ee 
where the boundary layer of thickness $\delta$ is defined by 
$$ \Omega_{\delta}:=\{{\bf x}\in \Omega: d({\bf x},\partial\Omega)<\delta\}.$$
This is the famous condition shown by Kato \cite{kato1984remarks} to be equivalent to strong $L^2$ convergence of the Navier-Stokes solution to a strong Euler solution, when the initial data 
also converge strongly and as long as the strong Euler solution exists. 
Since such strong limits of 
Navier-Stokes solutions are admissible weak Euler solutions, Kato's theorem can be 
considered one of the earliest results of weak-strong uniqueness type. 
\black{A more recent work of Kelliher \cite{kelliher2017observations} 
has comprehensively discussed
these various conditions for inviscid limits of closed flows in spatially bounded domains 
and shown, in particular, that the condition \eqref{BTcond} can be weakened to 
\be \lim_{\nu\to 0} \int_0^T \int_{\partial\Omega} 
{\boldsymbol \tau}_w^\nu\bdot {\bf U}\, dA\, dt= 0,  \lb{Kcond} \ee 
where ${\bf U}$ is an assumed strong Euler solution.}
Another class of conditions 
sufficient for weak-strong uniqueness involve some continuity of velocity for the admissible weak
Euler solution near the boundary. This was first proved by \cite{bardos2014non} who required that 
the velocity field be H\"older continuous in some neighborhood $\Omega_\epsilon$
with $\epsilon>0$ and then proved by \cite{wiedemann2018weak} assuming only continuity in $\Omega_\epsilon$. 
Recently, we have shown \cite{quan2025weak} that even less stringent conditions 
on the weak Euler solution are sufficient for weak-strong uniqueness within the general class
of admissible weak Euler solutions, namely:
\be \int_0^T \|\bu(\cdot,t)\|^2_{L^\infty(\Omega_\epsilon)} dt <\infty, 
\lb{DNcond1} \ee
for some $\epsilon>0,$ or near-wall boundedness, and further
\be \lim_{\delta\to0}\int_0^T \|\bn\bdot\bu(\cdot,t)\|^2_{L^\infty(\Omega_\delta)}dt = 0. \lb{DNcond2} \ee 
Note that 
$\|\bu\|_{L^\infty(\Omega)}:={\rm ess.sup}_{\mathbf{x}\in\Omega}|\mathbf{u}(\mathbf{x})|,$
where the ``essential supremum'' is the least upper bound of values $U$ such that the set
$\{\bx\in\Omega:\, |\bu(\bx)|>U\}$ has positive measure. 
The latter condition \eqref{DNcond2} can be interpreted as uniform continuity at the boundary 
of the normal velocity component, since $\left.\bu\bdot\bn\right|_{\partial\Omega}=0.$ We refer to 
\eqref{DNcond1}-\eqref{DNcond2} as the ``Drivas-Nguyen conditions'' since they were first 
employed in \cite{drivas2018nguyen} to study anomalous energy dissipation in wall-bounded flow. 

Although these conditional weak-strong uniqueness results have previously been interpreted
as identifying additional ``admissibility criteria'' for weak Euler solutions \cite{wiedemann2018weak},
we argue below that weak-strong uniqueness probably fails for the physical weak solutions and thus
all of these apparently modest conditions must be violated. A main example that motivates our claim 
is the d'Alembert potential Euler flow, which we next discuss. 

\section{Proof Sketch for D'Alembert Flow}\lb{sec:proof} 

To our knowledge, no prior theorem on conditional weak-strong uniqueness has been applicable to the d'Alembert potential Euler solution, until we recently established such a result \cite{quan2025weak}. The theorems of \cite{bardos2013mathematics} on inviscid limits apply to flow domains $\Omega$ of unbounded extent but assume that the smooth Euler solution has finite energy, which is untrue of the d'Alembert flow with a 
non-vanishing velocity $\bV(t)$ at infinity. Our result in \cite{quan2025weak} is likewise proved for inviscid limits $\bu=\lim_{\nu\to 0}\bu^\nu,$ where the initial data $\bu^\nu_0$ for the Navier-Stokes solution $\bu^\nu$ converge strongly 
in $L^2$ to the potential flow $\bu_\phi(\cdot,0)$ with velocity $\bV(0)$ at infinity. As we 
discuss below, this assumption is true both for the case of impulsive acceleration and for gradual 
acceleration from rest. We assume finite kinetic energy (spatial $L^2$ norm) only for the 
rotational fluid velocity $\bu_\omega:=\bu-\bu_\phi,$ which is realistic because this field 
is dipolar at infinity \cite{eyink2021josephson}. The main technical tool that we employ is the 
Josephson-Anderson relation for drag on the body \cite{eyink2021josephson}, which appears as a 
source term in the balance equation for kinetic energy of rotational fluid motions in the wake:
$$ E_\omega(t):=\frac{1}{2}\|\bu_\omega(\cdot,t)\|_2^2=\frac{1}{2}\|\bu(\cdot,t)-\bu_\phi(\cdot,t)\|_2^2.$$
We thus use a version of a standard ``relative energy'' argument to prove weak-strong uniqueness
\cite{wiedemann2018weak}. In this context, our sufficient condition for weak-strong uniqueness 
is \black{that 
$\int_{\partial \Omega} {\boldsymbol\tau}_w{\boldsymbol\cdot}{\bf u}_\phi\, dA \equiv 0$ 
for all times, which is nearly the same condition \eqref{Kcond} invoked by Kelliher 
for closed flows in bounded domains.} 

Because the essence of the proof in \cite{quan2025weak} is quite simple, it is instructive
to sketch the main details here. The starting point of the proof is the global balance for the 
rotational kinetic energy in the closed domain $\bar{\Omega}\times [0,T]$ so that for a.e. $\tau\in (0,T)$
\bea  E_{\omega}(\tau)&=& 
   E_{\omega}(0) -\int_0^{\tau} \int_\Omega Q \, dV\, dt+
\int_0^{\tau}\int_{\partial\Omega}
{\bf u}_\phi{\boldsymbol\cdot}{\boldsymbol\tau}_w \, dA\, dt \cr
   &&  \quad -\int_0^{\tau}\int_{\Omega}{\boldsymbol\nabla}{\bf u}_{\phi}\,{\boldsymbol :} \, 
    {\bf u}_{\omega}\otimes{\bf u}_{\omega}\,dV\,dt
\lb{rotbal} \eea 
Here $Q=\lim_{\nu\to 0} \nu|\bomega^\nu|^2\geq 0$ is the anomalous energy dissipation and 
the last two terms are the inviscid limit of the Josephson-Anderson relation for power dissipated
by drag, which appears as a source term  in the rotational energy balance. 
Much of the technical work of the proof involves the justification of the energy balance equation \eqref{rotbal}, by smearing a corresponding local balance equation with a test function $\varphi$ 
which is a smoothed version of the characteristic function $\chi_{(-\delta,\tau]}\chi_{B_R}$ 
\black{where $B_R$ is the ball of radius $R$ centered at the origin}, restricted to $\bar{\Omega}\times [0,T)$ 
and then taking the limit $R\to\infty.$ 

Assuming that ${\boldsymbol\tau}_w{\boldsymbol\cdot}{\bf u}_\phi\equiv 0$, then since $Q\geq 0$ 
\bea && E_{\omega}(\tau)\leq
   E_{\omega}(0) 
    -\int_0^{\tau}\int_{\Omega}{\boldsymbol\nabla}{\bf u}_{\phi}{\boldsymbol :} 
    {\bf u}_{\omega}\otimes{\bf u}_{\omega}\,dV\,dt \cr 
    &&\leq 
   E_{\omega}(0) 
    +C\int_0^{\tau} \|{\boldsymbol\nabla}{\bf u}_\phi(\cdot,t)\|_{L^\infty(\Omega)}E_\omega(t)\, dt, 
    \lb{main_ineq} \eea
where the last line follows by Cauchy-Schwartz inequality. Finally, by \eqref{main_ineq} and the Gronwall inequality, one obtains for a.e. $\tau\in (0,T)$ 
\be E_{\omega}(\tau) \le E_{\omega}(0)
\exp\left(C\int_0^{\tau}\|\grad{\bf u}_{\phi}(\cdot,t)\|_{L^{\infty}(\Omega)}dt\right). \lb{wsuniq} \ee
Thus, when ${\bf u}_0={\bf u}_\phi(\cdot,0)$ so $E_\omega(0)=0,$
then it follows that 
$${\bf u}(\cdot,\tau)={\bf u}_\phi(\cdot,\tau), \qquad {\rm a.e.}\ \tau\in (0,T).$$
This argument shows that any strong-$L^2$ inviscid limit $\bu$ with initial data $\bu_0=\bu_\phi(\cdot,0)$
must in fact coincide with $\bu_\phi$ for a.e. times $\tau\in [0,T]$. 
The applicability of this result to a smoothly accelerated body with $\bV(0)=\bzed$ is straightforward,
assuming the validity of the incompressible Navier-Stokes model, since $\bu^\nu=\bu_\phi\equiv\bzed.$ 
The relevance to the impulsively accelerated body requires more detailed justification because 
of the singular nature of that problem. It is worthwhile to discuss here since it raises some physics issues which will be important later. 

Our derivation of the inviscid Josephson-Anderson relation \cite{quan2024onsager} and 
rigorous proof of weak-strong uniqueness for the d'Alembert potential Euler solution 
\cite{quan2025weak} both assume existence of a strong Navier-Stokes solution 
on the domain $\bar{\Omega}\times [0,T)$ for any viscosity $\nu>0.$ However, this 
assumption cannot be true up to time $t=0$ for the initial data $\bu^\nu(\cdot,0)=\bu_\phi,$
because ${\boldsymbol \Gamma}:=\bn\btimes \bu_\phi\neq \bzed$ and the first
compatibility condition for a continuous solution is thus violated 
\cite{temam2006suitable}. However, the Navier-Stokes 
solution for this initial data can be expected to be smooth on the time-interval
$[t_0,T]$ for any time $t_0>0$ \cite{temam2006suitable}.
Indeed, for the Prandtl problem of an impulsively accelerated cylinder, a formal 
Navier-Stokes solution has been 
obtained for $Re\gg 1$ at short dimensionless times $t\sim \alpha/Re,$ with $\alpha$ fixed 
but arbitrary \cite{bar1975initial}. The non-dimensionalization here employs outer units, 
with lengths normalized by the cylinder radius $R,$ velocities normalized by the flow velicity
$V$ at infinity, and times normalized by $R/V.$ The formal solution is obtained by a third-order matched 
asymptotic expansion in which the outer potential-flow solution to leading order is the 
d'Alembert solution $\bu_\phi$ and the inner solution 
describes a viscous boundary layer of thickness $\sim(t/Re)^{1/2}$ at the cylinder suface. 
Although this formal solution has never been rigorously derived, to our knowledge, it agrees
well with high-resolution numerical simulations of the Prandtl problem at early times
\cite{koumoutsakos1995high,chatzimanolakis2022vortex}. Since the boundary layer of thickness
$\sim(t/Re)^{1/2}$ corresponds to $\bu_\omega^{Re}(t)$ with vanishingly small kinetic energy 
as $Re\to\infty,$ we may consider instead the Navier-Stokes solution on the time interval 
$[t_0,T]$ with $t_0=\alpha/Re$ for some fixed $\alpha$ and it remains true that 
$\lim_{Re\to\infty}\bu^{Re}(t_0)=\bu_\phi$ strongly in $L^2.$

The above mathematical explanation nevertheless suffers from a physical defect, 
because the formal asymptotic solution has a skin friction diverging as $Re^{-1/2}t^{-1/2}$ 
at $t=0,$ due to the initial singular vortex sheet at the body surface \cite{bar1975initial}. 
This solution for impulsive acceleration of the body is obviously experimentally unrealizable. 
If the acceleration of the body really occurs on a time scale of order the mean-free 
collision time of the molecular fluid or on an even shorter time scale, then the problem can no 
longer be described accurately by Navier-Stokes equation with stick boundary conditions.
A more detailed analysis by methods of kinetic theory \cite{reddy1967rayleigh} suggests 
that one may still obtain a uniformly accurate solution by Navier-Stokes equations, if one 
replaces the stick boundary conditions with Navier-slip conditions for a slip length of order 
the molecular mean-free-path length $\lambda_{mfp}$ and if the initial data are modified 
by a ``kinetic boundary layer'' or Knudsen layer with thickness also of order $\lambda_{mfp}$. The latter 
layer removes the divergence in the skin friction at $t=0$, as physically required. 
Note that this boundary layer in dimensionless outer units has thickness $\lambda_{mfp}/R\sim Ma/Re,$
with $Ma=c_s/V$ the Mach number, and the kinetic description modifies the  previous asymptotic 
Navier-Stokes solution $\bu^{Re}(t)$ at very short times $t\sim Ma^2/Re.$ We used here 
the standard estimate from kinetic theory for kinematic viscosity
$\nu\sim \lambda_{mfp}c_s,$ with $\lambda_{mfp}$ the mean-free-path length and 
with $c_s$ the sound speed. The drag coefficient 
no longer diverges at $t=0$ but instead assumes a large value $\sim Ma^{-1}$ \cite{reddy1967rayleigh}. 
The proof of weak-strong uniqueness for the d'Alembert potential flow given in \cite{quan2025weak} works 
with Navier-slip b.c. as long as the slip length in dimensionless outer units vanishes as $Re\to\infty.$ See Appendix \ref{app:slip}. 
Thus, our mathematical results still apply in this kinetic reformulation. 

Some doubts remain, however, whether our analysis applies to a physical realization of 
impulsive acceleration in a laboratory experiment. The kinetic description of a gas
by the Boltzmann equation in the low-density limit is now believed to 
be incomplete, missing stochastic effects of molecular fluctuations 
\cite{spohn1981fluctuations,bouchet2020boltzmann,sawant2021kinetic}. 
Such fluctuation effects do not necessarily vitiate Onsager's theoretical description of 
high Reynolds number flows by weak Euler solutions \cite{eyink2024space}. However, 
to our knowledge, weak-strong uniqueness results have not yet been proved for inviscid limits
of models incorporating fluctuations. Thus, it seems advisable to focus theoretical 
treatment and future empirical investigation on the more regular (and commonplace) problem 
of gradually and smoothly accelerated bodies. Rather surprisingly, however, we shall find that even 
the problem of smooth acceleration may encounter such difficulties, because breakdown 
of weak-strong uniqueness requires such extreme events that the validity of a macroscopic
hydrodynamic description again is threatened. See Section \ref{sec:break}.  

\section{Two Alternative Scenarios}\lb{sec:alt} 

Assuming validity of a hydrodynamic description by Navier-Stokes, our mathematical results 
lead to two distinct alternatives, both of which might be argued to be consistent with empirical 
observations of turbulent drag on solid bodies. We discuss these two alternatives in turn. 

\subsection{First Alternative:  ${\boldsymbol\tau}_w{\boldsymbol\cdot}{\bf u}_\phi\equiv 0$}

Under this condition, weak-strong uniqueness holds. As a consequence, for initial data strongly converging to pure potential, 
drag on a body vanishes over any fixed time-interval $[0,T]$ as $Re\to\infty;$ see \eqref{dragzero}. 
If the drag vanishes sufficiently slowly, however, then it might be very difficult to distinguish
empirically from drag strictly non-vanishing. The phenomenon of very slowly vanishing drag and dissipation  
has been termed a ``weak dissipative anomaly'' \cite{bedrossian2019sufficient}. Such weak anomalies
seem to occur physically in a number of wall-bounded flows, such as circular pipe flows with 
perfectly smooth walls (see \cite{eyink2024onsager} for a summary). \black{Convergence to a smooth 
potential Euler solution has been suggested to occur also in some bluff body flows, e.g. rapidly 
rotating cylinders where separation is argued to be suppressed \cite{glauert1957flow}, 
although recent high Reynolds number simulations do not seem to support that conclusion
\cite{aljure2015influence}.}
Strong anomalous dissipation and drag 
may furthermore occur experimentally if the initial data is not pure potential in the high-Reynolds number limit. 
Although the rotational flow from a thin viscous boundary layer vanishes in the energy norm 
as $Re\to\infty,$ there may in addition be small but non-vanishing vorticity in the incoming flow. This 
is similar to what occurs in so-called ``bypass transition'', where small levels of turbulent 
fluctuations in the background flow can trigger laminar-to-turbulent transition. In order to discuss 
both of the above possibilities more quantitatively, we can apply the analysis of the previous section. 

The first possibility of ``weak anomaly'' is best discussed by generalizing this analysis to 
finite Reynolds number $Re<\infty,$ or equivalently positive viscosity $\nu>0$. From the global 
energy balance for the rotational motions, one obtains
\bea E_{\omega}^\nu(\tau)&=& 
   E_{\omega}^\nu(0) +
\int_0^{\tau}\int_{\partial\Omega}
{\bf u}_\phi{\boldsymbol\cdot}{\boldsymbol\tau}_w^\nu \, dA\, dt
        -\int_0^{\tau} \int_\Omega Q^\nu \, dV\, dt \cr
&&\quad     -\int_0^{\tau}\int_{\Omega}{\boldsymbol\nabla}{\bf u}_{\phi}\,{\boldsymbol :} \, 
    {\bf u}^\nu_{\omega}\otimes{\bf u}^\nu_{\omega}\,dV\,dt \cr 
&\leq &
   E_{\omega}^\nu(0)  +
\int_0^{\tau}\int_{\partial\Omega}
{\bf u}_\phi{\boldsymbol\cdot}{\boldsymbol\tau}_w^\nu \, dA\, dt \cr
&&\quad    +C\int_0^{\tau} \|{\boldsymbol\nabla}{\bf u}_\phi(\cdot,t)\|_{L^\infty(\Omega)}E_\omega^\nu(t)\, dt \eea 
by using $Q^\nu=\nu |\bomega^\nu|^2\geq 0$ and Cauchy-Schwartz inequality. Then by Gronwall inequality, we obtain
\bea 
&&E_{\omega}^\nu(\tau) \le \left(E_{\omega}^\nu(0) +
\int_0^{\tau}\int_{\partial\Omega}
{\bf u}_\phi{\boldsymbol\cdot}{\boldsymbol\tau}_w^\nu \, dA\, dt\right) \cr
&& \quad \qquad \times \exp\left(C\int_0^{\tau}\|\grad{\bf u}_{\phi}(\cdot,t)\|_{L^{\infty}(\Omega)}dt\right)
\eea 
Even though the prefactor of the exponential vanishes in the limit $Re\to\infty$ with $\tau$ fixed,
the upper bound does not generally vanish for fixed $Re\gg 1$ as $\tau\to\infty.$ Consider 
the case where the initial rotational flow arises entirely from a viscous boundary layer 
of vanishingly small thickness $\delta/D\sim Re^{-\alpha}$ for $Re\gg 1$ with $\alpha>0.$ In that case, 
$E_\omega^\nu(0)\rightarrow 0$ as $\nu\to 0$ and, by our main assumption in this section, also
${\boldsymbol\tau}_w^\nu{\boldsymbol\cdot}{\bf u}_\phi\to 0.$ However, when $\alpha<1$ (as for a 
Prandtl boundary layer with $\alpha=1/2$), then the Reynolds number of the boundary layer itself
will be very large for $Re_\delta=U\delta/\nu=Re^{1-\alpha}\gg 1$ and thus prey to various instabilities.
The successive instabilities of Prandtl layers and subsequent thinner sublayers have been studied both by 
fluid mechanicians \cite{cassel2014unsteady} and by mathematicians \cite{grenier2018sublayer,grenier2019instability}.
Such instabilities provide a mechanism for generation and growth of rotational flow, so that 
$$E_\omega^\nu(\tau)\gg E_\omega^\nu(0), \qquad \tau \rightarrow \infty, \quad \nu\ {\rm fixed}. $$
Even in very quiet flow without tiny external perturbations, intrinsic thermal noise 
might trigger instability and flow separation \cite{mcmullen2024hydrodynamic}.

It is important in this context to distinguish between two different notions of stability.
On the one hand, there is the concept of stability in mechanics or dynamical systems, according 
to which a solution is ``stable'' if small perturbations do not grow or even decay in magnitude.
In the opposite case where infinitesimal perturbations grow in magnitude, the solution is called 
{\it dynamically unstable}. On the other hand, another notion of stability of solutions 
in applied mathematics is ``well-posedness'', which holds when existence, uniqueness 
and continuity in the data (initial conditions, equations of motion, etc.) are all
guaranteed. This property is sometimes called ``Hadamard stability'' after the mathematician
who first codified the concept \cite{hadamard1902problemes}. In the opposite case, the solution
is called {\it ill-posed} or {\it Hadamard unstable}. What must be emphasized is that 
Hadamard stability is generally a much weaker requirement than dynamical stability and 
it is perfectly consistent with exponential growth of small perturbations. For example,
even smooth chaotic dynamical systems in which every solution exhibits exponential sensitivity 
to initial data are well-posed in the sense of Hadamard. 

{\it Weak-strong uniqueness is precisely a statement of well-posedness of the classical smooth
Euler solution, i.e. its Hadamard stability, even within a much larger class of “admissable, 
generalized Euler solutions”.} This point is made very clearly by the basic inequality 
\eqref{wsuniq}. This result implies that any ``viscosity solution'' of Euler, $\bu,$ obtained by a 
strong inviscid limit, must coincide with the smooth potential solution
of d'Alembert, $\bu_\phi,$ if $\bu(\cdot,0)=\bu_\phi(\cdot,0)$ or $E_\omega(0)=0.$ In addition
to this uniqueness statement, one can infer also from \eqref{wsuniq} continuity in the initial data:
any ``viscosity solution'' $\bu$ can be made to agree arbitrarily closely with $\bu_\phi$ over any 
finite time interval $[0,T],$ if $\|\bu(\cdot,0)-\bu_\phi(\cdot,0)\|_2\ll 1$
or $E_\omega(0)\ll 1.$ As we have just discussed, however, this ``Hadamard stability''
of the d'Alembert solution $\bu_\phi$ is perfectly compatible with its dynamical instability. 
This fact is directly relevant to the situation where the initial data $\bu^\nu_0$ for Navier-Stokes do 
not converge strongly in $L^2$ to $\bu_\phi(\cdot,0),$ because turbulent fluctuations with
some small but non-vanishing energy are superimposed on the potential flow. At long times the 
viscosity solution $\bu$ with such initial data can depart far from the d'Alembert solution $\bu_\phi$ 
and can exhibit a strong dissipative anomaly. 

These conclusions have implications for some interesting proposals of Hoffman \& Johnson 
on the solution of the d'Alembert paradox \cite{hoffman2006simulation,hoffman2007computational,hoffman2010resolution}.
In agreement with the earlier ideas of Onsager, they explain non-vanishing or anomalous drag via 
``turbulent Euler solutions'', which they have attempted to calculate from numerical finite-element 
schemes. They summarize their proposed solution as follows: 
\begin{quotation}
``We have presented a resolution of d’Alembert’s Paradox based on analytical and 
computational evidence that a potential solution with zero drag is illposed as a solution of the Euler 
equations, and under perturbations develops into a wellposed turbulent solution with substantial drag 
in accordance with observations'' \cite{hoffman2010resolution}. 
\end{quotation}
Recognizing that such weak Euler solutions may be non-unique
even for exactly specified initial data $\bu_0,$ the authors suggest an interesting concept of 
``output well-posedness'' according to which some space-time averaged outputs (such as mean drag)
may be unique and continuous in the inital data, even though individual Euler solutions are not.
These proposals are generally consistent with our rigorous mathematical analysis, with just one 
important exception: the d'Alembert potential flow can be dynamically unstable but it cannot be,
as claimed by \cite{hoffman2010resolution}, ``ill-posed as a solution of the Euler equations,'' 
at least not when weak-strong uniqueness holds. As we discuss in the following section, and at length 
in Section \ref{sec:ext}, ill-posedness of the d'Alembert potential flow in fact requires very 
extreme wall events with quite striking experimental signatures.

\subsection{Second Alternative: ${\boldsymbol\tau}_w{\boldsymbol\cdot}{\bf u}_\phi\not\equiv 0$}

When the condition ${\boldsymbol\tau}_w{\boldsymbol\cdot}{\bf u}_\phi= 0$ is not valid identically
in spacetime (in the sense of distributions), then weak-strong uniqueness may not hold for the 
d'Alembert potential flow within the class 
of strong inviscid limits. In particular, the potential Euler solution ${\bf u}_\phi$ may co-exist 
with other weak Euler solutions ${\bf u}$ obtained via inviscid limits, with identically the same initial data, 
$\bu(\cdot,0)={\bf u}_\phi(\cdot,0),$ but with non-vanishing vorticity and drag! This scenario recalls 
the result for plane-parallel channel geometry 
of \cite{vasseur2023boundary} that plug flow coexists with infinitely-many, admissible weak Euler solutions 
exhibiting separation. As we discuss in the next Section \ref{sec:ext}, available evidence suggests 
that the conditions for weak-strong uniqueness in fact hold in channel flows. On the other hand,
we shall argue that breakdown of weak-strong uniqueness is a very plausible possibility for flows around 
solid bodies, permitting coexistence of the smooth d'Alembert flow and dissipative Euler flows with 
separation and drag. We discuss next the extreme flow events that are required for such breakdown to occur.  

\section{Extreme Wall Events}\lb{sec:ext} 

We have already discussed in Section \ref{sec:fail} three rather modest-looking conditions that suffice 
for weak-strong uniqueness. Within the class of weak-* limits of Navier-Stokes solutions with bounded energy
in general domains, \cite{bardos2013mathematics} have shown that the condition \eqref{BTcond} of vanishing 
skin-friction suffices for weak-strong uniqueness of any smooth Euler solution. Furthermore, 
\cite{bardos2013mathematics} have shown that the condition \eqref{Kcond} of vanishing energy 
dissipation in a ``Kato layer" implies the previous condition \eqref{BTcond}. In fact, the third 
set of conditions \eqref{DNcond1},\eqref{DNcond2} on continuity of normal velocity at the wall 
implies also the vanishing skin-friction condition \eqref{BTcond} for strong inviscid limits 
of Navier-Stokes solutions. The proof in \cite{quan2022inertial} is based on the analogy between 
skin friction and viscous energy dissipation, on the one hand, and energy cascade through scales 
and momentum cascade through space, on the other hand \cite{jimenez2012cascades}. Thus,
non-vanishing skin friction $\btau\not\equiv\bzed$ can be understood as a ``strong momentum 
anomaly'' and the continuity conditions \eqref{DNcond1},\eqref{DNcond2} are analogous to the 
Onsager condition on velocity H\"older exponent $h>1/3,$ which forbids an energy to small 
scales. The analogous idea of the proof in \cite{quan2022inertial} is that non-vanishing 
velocity toward the wall, at any positive distance, is required to carry momentum to the viscous sink 
at the wall in the infinite Reynolds number limit. 
Thus, a strong momentum anomaly is possible only if the normal velocity 
$v:={\bf n}{\boldsymbol\cdot}{\bf u}$ is \black{is not uniformly continuous} at the wall. 

Somewhat surprisingly, the conditions for a momentum 
anomaly and breakdown of weak-strong uniqueness are much more severe than those needed 
for energy cascade, requiring essentially a shock-like \black{``discontinuity''} at the wall, or an $h=0$ 
H\"older singularity! Interestingly, this possibility seems to have been anticipated 
by G. I. Taylor as early as 1915, who wrote the following in discussing interactions 
of atmospheric turbulence with the solid ground: 
\begin{quotation}
\noindent 
``...a very large amount of momentum is communicated by means of eddies from the atmosphere to the ground. 
This momentum must ultimately pass from the eddies to the ground by means of the almost infinitesimal 
viscosity of the air. The actual value of the viscosity of the air does not affect the rate at which momentum is
communicated to the ground, although it is the agent by means of which the transference is effected.    

\noindent ...

\noindent The finite loss of momentum at the walls due to an infinitesimal viscosity may be compared 
with the finite loss of energy due to an infinitesimal viscosity at a surface of discontinuity in a gas.'' 
\cite{taylor1915eddy}
\end{quotation}
The Kato condition \eqref{Kcond} likewise involves an extreme situation of non-vanishing viscous dissipation in a shock-like layer at a wall. Kato-type boundary layers of thickness $\propto 1/Re,$ 
it is worth noting, were proposed already in 1923 by Burgers as a mechanism to produce anomalous drag \cite{burgers1923resistance}\footnote{Burgers' discussion was longer than 
Taylor's \cite{taylor1915eddy} but his arguments, as he admitted himself, were 
``in many respects vague''. He proceeded by deriving integral mean balances for momentum and 
energy and by investigating the parameter values to maximize drag. He correctly concluded
that a boundary layer with thickness $\sim 1/Re$ would produce skin friction non-vanishing 
as $Re\to \infty.$ Unfortunately, this argument leads to non-vanishing drag coefficient 
$C$ for smooth walls, whereas Burgers had noted himself that only ``for rough-walled
tubes $C$ is approximately independent'' of $Re$.}. 
The corresponding rigorous conditions for weak-strong uniqueness,
although originally derived for internal flows in bounded domains, 
apply also to external flows around bodies: see Section \ref{sec:proof}.

The severe events required to violate weak-strong uniqueness possess very striking signatures
that should be observable empirically, if they exist, both in numerical simulations and in laboratory 
experiments. There is already very active current investigation of extreme events in wall-bounded 
turbulence. However, the events required to violate weak-strong uniqueness are far more 
violent than those reported in the prior literature, with one prominent exception
\cite{nguyenvanyen2018energy}. This relative lack
of evidence may be due to the dominant focus of such studies on the ``canonical wall-bounded flows'':
plane-parallel channel flow \cite{lenaers2012rare,srinivas2024parametrizing}, 
circular pipe flow \cite{guerrero2020extreme,fei2024extreme}, 
and the flat-plate boundary layer \cite{pan2018extremely}. Although these flows are often 
considered to capture the essence of turbulent-solid surface interactions in the simplest 
geometry, in fact these flows exhibit a number of very atypical features, 
as a direct consequence of their simplicity. For example, none of these canonical flows show any evidence 
for a strong energy dissipation anomaly, in contrast to common wake flows past solid bodies
or internal flows with hydraulically rough walls \cite{eyink2024onsager}. We summarize 
below current observations on extreme events, first for channels as representative
of ``canonical flows'' and then for the single flow that, to our knowledge, provides positive evidence. 

\subsection{Channel Flow}\lb{sec:channel} 

The relevant quantities for weak-strong uniqueness are skin friction, wall-normal velocity 
and near-wall viscous dissipation, which we discuss in turn. 

\subsubsection{Skin Friction}

The condition ${\boldsymbol\tau}_w\not\equiv \bzed$ may seem improbable, given the very common 
observation that time-average skin friction vanishes in the high Reynolds limit, 
$\bar{{\boldsymbol\tau}}_w^\nu\to {\bf 0}$ as $\nu\to 0.$ This observation even holds 
when drag is non-vanishing, as for bluff bodies \cite{achenbach1972experiments} and  
rough pipes \cite{busse2017reynolds}, because the asymptotic net drag in those instances is 
apparently supplied entirely by pressure forces (form drag). However, in order to violate 
weak-strong uniqueness it is enough that ${\boldsymbol\tau}_w$ not vanish over some finite 
region of space-time and it need not be true that any non-zero fraction of the asymptotic 
drag arises from skin friction. Thus, rather than the mean value, it is more appropriate 
to consider the entire distribution of values over the wall at long times. 

\begin{figure}
    \centering
    \includegraphics[width=\linewidth]{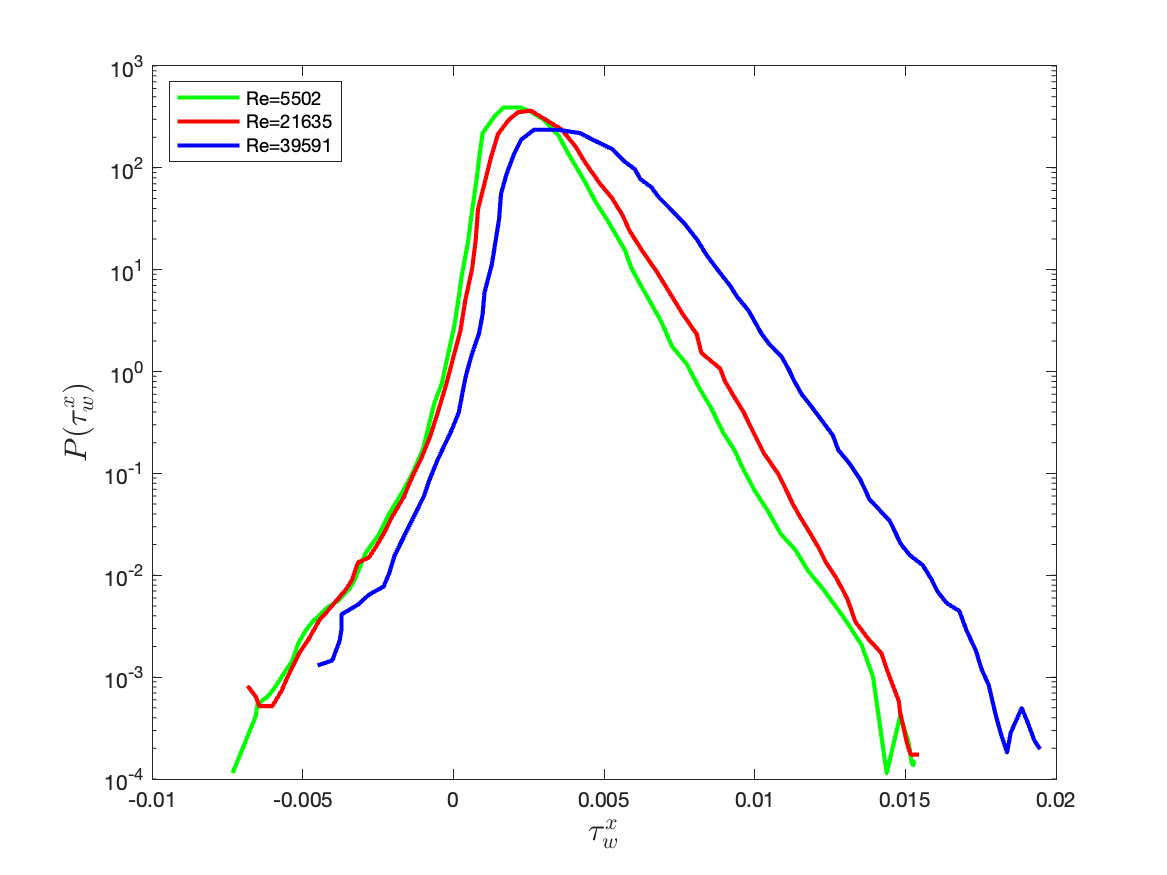}
    \caption{Plot of the PDF of the streamwise component of skin friction in turbulent channel
    flow from the numerical simulation data in \cite{lenaers2012rare}, but scaled in outer units.}
    \label{fig:skinzero}
\end{figure}

The probability distribution of wall shear stress has been a focus for previous numerical studies
\cite{lenaers2012rare,srinivas2024parametrizing}, in particular the streamwise component 
$\tau_w^x$ which contributes to drag. Note that it is also non-vanishing of the streamwise 
component which is relevant for possible physical violation of weak-strong uniqueness 
for plug-flow in a channel, with $\bu_\phi= U\hat{\bx}$ \black{a spacetime constant velocity in the 
streamwise or $x$-direction.} 
In fact, since all of the theorems 
about infinite-$Re$ limit cited in this paper require non-dimensionalization of flow variables
in outer units, what is relevant is $\tau_w^x$ scaled by $U^2.$ Since the results in 
\cite{lenaers2012rare,srinivas2024parametrizing} are instead scaled by 
$\bar{{\tau}}_w^x=u_\tau^2,$ the friction-velocity squared, we have 
scanned the data in Figure 2 of \cite{lenaers2012rare} and rescaled the results 
by the factor $u_\tau^2/U^2.$ The parameterization of the Prandtl-K\'arm\'an drag law
from equation (12) of \cite{zanoun2021wall} was used to determine 
$Re=U(2h)/\nu$ from $Re_\tau=u_\tau h/\nu$ and hence the ratio $U/u_\tau.$ 
The results for the probability density function of $\tau_w^x$ in outer units
are plotted in Figure \ref{fig:skinzero}, at three values of the Reynolds number $Re.$

The trends with increasing Reynolds number are clear. The mean values $\bar{{\tau}}_w^x$
are decreasing very slowly, consistent with the logarithmic decay predicted by the 
Prandtl-K\'arm\'an law. Simultaneously, the widths of the probability distributions 
are shrinking slowly also with increasing $Re.$ This decrease in widths is a consequence of our scaling 
in outer units, whereas prior observations of the probability distribution of skin friction 
in inner units revealed increasing variances and increasing far tails as Reynolds numbers rose \cite{lenaers2012rare}. 
Although the convergence suggested by our Figure \ref{fig:skinzero} is quite slow, the results
are consistent with $\tau_w^x\to 0$ as \black{$Re\to \infty.$} According to this empirical data, 
weak-strong uniqueness is likely to hold for plug flow through the channel, 
within the class of inviscid limits. 

\subsubsection{Wall-Normal Velocity}

The Drivas-Nguyen conditions for weak-strong uniqueness involve, in particular, 
extreme values of the wall-normal velocity (essential suprema) which are non-vanishing 
approaching the wall. Extremes of the wall-normal velocity have previously been studied 
in channel flow, for example, see \cite{lenaers2012rare}, Figure 5, for probability 
distributions of the wall-normal velocity. However, those distributions are for 
wall-normal velocities scaled by their standard deviation and involve positions only 
in the viscous sublayer and buffer layer. Instead, the Drivas-Nguyen condition involves $v$-extremes 
in the inertial layer where direct viscous effects are entirely absent. In dimensional terms,
the Drivas-Nguyen condition on wall-normal velocity is that 
\be \lim_{\delta\to 0}{{1}\over{T}}\int_0^T\left(\max_{x,z,\, 0<y<\delta \,H}|v^\nu({\bf x},t)|\right)^2 dt \geq (\epsilon U)^2, \lb{DNcond3} \ee 
for some $\epsilon>0.$ Thus, the maximum of $v^\nu/U$ for distances $y<\delta\cdot H$ must remain $\geq\epsilon$ for some 
$\epsilon>0$ as $\delta\to 0$ through the inertial layer.  
The maximization can presumably be restricted to the inertial interval $nu/u_\tau\ll y<\delta\cdot H$ 
(in inner units, $y^+\gg 1$), 
since the largest values should occur there. There have been some previous studies of extreme 
values in the logarithmic layer of turbulent channel flow, but for viscous dissipation rather 
than velocities \cite{hack2021extreme}. There is some indication from Figure 6 of \cite{lenaers2012rare}
that extremes of wall-normal velocity become highly improbable at distances $y^+=yu_\tau/\nu>10,$
if an ``extreme value'' is defined as there to be a value greater than 10 standard deviations from 
the mean. However, note that the condition \eqref{DNcond3} requires instead velocities only 
some small fraction $\epsilon>0$ of $U,$ but arbitrarily close to the wall. We are aware 
of no direct evidence for this condition in channel flow, but it is not implausible. 

\subsubsection{Dissipation in a Viscous Near-Wall Layer}

The Kato condition \eqref{Kcond} for channel flow is just the requirement that the net energy dissipation 
integrated over the viscous sublayer and buffer layer should vanish relative to $U^3/H$ as $Re\to \infty.$
Although there have been a great many studies of viscous dissipation in turbulent channel flows, we are not 
aware of any study of this particular issue. The prediction of the Prandtl-K\'arm\'an theory is that dissipation 
in the buffer layer and viscous sublayer should scale as $\sim u_\tau^3/H,$ smaller than the total dissipation 
in the log-layer by a factor of about $\log Re_\tau.$ Both, however, tend to zero relative to $U^3/H$ as $Re\to \infty.$ These considerations suggest that the Kato condition should be satisfied and indeed 
convergence to plug flow has been argued to occur in smooth-walled, plane-parallel channel flows \cite{cantwell2019universal}. 

If so, then the counterexamples of \cite{vasseur2023boundary} to weak-strong
uniqueness in channel flows are of mathematical relevance only and would not appear in the inviscid limit.
Note, however, that all of the observations that we have discussed are for statistically stationary,
fully-developed turbulent channel flow. Weak-strong uniqueness involves instead the initial-value problem starting 
with plug flow, which may be achieved, for example, by impulsive acceleration of the channel walls to velocity 
$-U$. We are not aware of any relevant empirical studies in this context. 

\subsubsection{Vortex-Induced Separation}

Although most available observations on channel flow do not involve the initial transient regime 
and are thus not directly relevant to the issue of weak-strong uniqueness, it is nevertheless of 
some interest to understand what fluid mechanical events lead to the most extreme near-wall
events in fully-developed channel flow. The events which produced large wall-normal velocities 
(both positive and negative) were visualized in Figure 9 of \cite{lenaers2012rare} and as a composite 
image obtained by conditional averaging in figure 10 of of \cite{lenaers2012rare}. These results, confirmed 
by subsequent studies, show that extreme wall-normal velocity events in the viscous sublayer and buffer layer 
of turbulent channel flow with smooth walls have the general characteristics of vortex-induced separation \cite{doligalski1994vortex}. These extreme events are triggered by strong quasi-streamwise vortices which 
approach near the wall and induce very strong motions both toward and away from the wall. Extreme values of
wall-stress \cite{sheng2009buffer} and energy dissipation \cite{hack2021extreme} are induced by 
the same mechanism. 

It is noteworthy that vortex-induced separation is associated with blow-up of solutions of 
the Prandtl boundary-layer equations, leading to diverging wall-normal velocities \cite{peridier1991vortex}. 
This type of singularity was first identified numerically by Dommelen \& Shen in the 
Prandtl problem of an impulsively accelerated cylinder \cite{vandommelen1980spontaneous,vandommelen1982genesis}
and the blow-up has since been rigorously established \cite{kukavica2017van,collot2021singularities}. 
The boundary-layer equations cannot explain quantitatively all of the features of such extreme events,
but blow-up of the Prandtl solutions is probably a necessary antecedent. 

\subsection{Smooth Vortices Impinging on a Wall}\lb{sec:dipole} 

Given the above observations, a promising situation in which to observe anomalous dissipation
and breakdown of weak-strong uniqueness is the problem of compact vortices in an otherwise 
quiescent flow impinging on a solid wall. The numerical study of the simplest such example,
a dipole pair of vortices impacting on a flat wall in two space dimensions, was pioneered by Orlandi 
\cite{orlandi1990vortex} and studied subsequently at higher resolution 
\cite{clercx2002dissipation,nguyenvanyen2018energy}. This example shares with the d'Alembert 
potential flow the very important feature that a smooth Euler solution exists globally in time,
in this case because of the restriction to two space dimensions. The smooth Euler solution corresponds 
here to the pair of vortices hitting the wall and then, under the influence
of their image vortices, propagating along the wall in opposite directions.

The latest numerical solutions \cite{nguyenvanyen2018energy}
of the Navier-Stokes equation with this same initial data at very 
high Reynolds numbers show quite different behavior, with the vortices rebounding from the wall
and inducing thereby a cascade of separation of vorticities of alternating signs. This
high-$Re$ solution is vividly compared with the smooth Euler solution in a supplementary movie of \cite{nguyenvanyen2018energy} at \url{https://doi.org/10.1017/jfm.2018.396}. The highest 
Reynolds number achieved by \cite{nguyenvanyen2018energy} was $Re=UL/\nu=123\,075,$ where
$U$ is the initial maximum velocity and $L$ is a measure of the radius of the initial vortices.
A special adaptive grid was used for the computations to assure that length scales of order 
$\sim 1/Re$ could be accurately resolved in the vicinity of the wall. This simulation provides
rather convincing {\it prima facie} evidence for co-existence of a smooth Euler 
solution and a quite distinct inviscid limit solution with the same initial data.

Most importantly, the paper \cite{nguyenvanyen2018energy} presented strong evidence also for the 
extreme events that are required in order to violate weak-strong uniqueness. Figure 12a of 
\cite{nguyenvanyen2018energy} plotted versus Reynolds number the maximum vorticity at the wall, 
which was found to scale as $Re^{1/2}$ before the blow-up of the Prandtl solution but 
as $Re^1$ a short time after the blow up. Since skin friction is related to wall vorticity 
by $\btau_w^\nu=\nu\bn\btimes\bomega^\nu$ for stick b.c., the above observation corresponds to $\btau_w^\nu\not\to\bzed$
as $\nu\to 0,$ at least pointwise. Furthermore, Figure 12b of \cite{nguyenvanyen2018energy} plotted 
also versus Reynolds number the enstrophy $\Omega=(1/2)\int_\Omega |\bomega^\nu|^2 dV,$
which was likewise found to scale as $Re^{1/2}$ before the Prandtl blow-up but 
as $Re^1$ afterward. Because total viscous energy dissipation can be written as $2\nu \Omega,$
the latter scaling implies a dissipative anomaly. No evidence was provided 
in \cite{nguyenvanyen2018energy} for violation of the Drivas-Nguyen conditions \eqref{DNcond1},
\eqref{DNcond2}, but the authors did note ``... a blow-up of the wall-normal velocity associated 
with an abrupt acceleration of fluid particles away from the wall'' (p.697). Although further 
confirmation would be desirable, the paper \cite{nguyenvanyen2018energy} presents to our knowledge
the most complete evidence for physical violation of weak-strong uniqueness
within the class of inviscid limits. 

\subsection{Breakdown of Deterministic Navier-Stokes?}\lb{sec:break}. 

An important feature of the conditions required to violate weak-strong uniqueness is that they are 
so extreme that they threaten the validity of a macroscopic hydrodynamic description. To explain this,
we may argue phenomenologically. Associated to local, instantaneous skin friction ${\boldsymbol \tau}_w^\nu({\bf x},t)$ on a body surface, one can introduce 
a fluctuating viscous length 
$$\delta_\nu({\bf x},t):={{\nu}\over{|{\boldsymbol \tau}_w^\nu({\bf x},t)|^{1/2}}}.$$
This length is analogous to the fluctuating dissipation scale considered in bulk turbulence 
\cite{paladin1987degrees,schumacher2007sub}. Note that a strong momentum anomaly corresponds 
to ${\boldsymbol \tau}_w^\nu({\bf x},t)\sim U^2,$ which yields the ``Kato length” 
$\delta_\nu({\bf x},t)/L\sim Re^{-1}.$
However, using again the estimate from kinetic theory that 
$\nu\sim \lambda_{mfp}c_s,$ then ${\boldsymbol \tau}_w^\nu({\bf x},t)\sim U^2$ gives also 
$$ \delta_\nu({\bf x},t)\sim \lambda_{mfp}\, Ma^{-1}, $$
where $Ma=U/c_s$ is the Mach number. This $\delta_\nu$ is only larger than $\lambda_{mfp}$ by $Ma^{-1}$
and thus dangerously close to length-scales at which no hydrodynamic description can be accurate.
Furthermore, thermal fluctuations become sizable already at lengths $\gg \lambda_{mfp}$
\cite{dezarate2006hydrodynamic,bandak2022dissipation,bell2022thermal}. Thus, a quantitatively correct 
description of such extreme wall events is probably not provided by deterministic Navier-Stokes 
equations but instead by some version of fluctuating hydrodynamics including fluid-solid friction effects \cite{reichelsdorfer2016foundations}. 

What is crucial to emphasize is that these tiny lengths may arise not only for the case of 
impulsive acceleration, which is obviously very singular, but even for apparently much more regular 
flows, for example, with gradually accelerated smooth bodies or smooth dipolar vortices impinging on a flat wall.
If the extreme events necessary to violate weak-strong uniqueness do not occur, then
inviscid limits with smooth initial data necessarily coincide with the smooth Euler solution
as long as that exists. A plausible mechanism to produce such extreme events is explosive 
separation of thin boundary-layers at extremely high Reynolds numbers. 

\section{Conclusions} 

According to the mathematical results reviewed in this work, there are two main scenarios 
for very high Reynolds-number fluid flows interacting with solid walls, depending upon which 
of the following two limits holds: (i) $\lim_{\nu\to 0}{\boldsymbol \tau}_w^\nu\equiv {\bf 0}$ or (ii) $\lim_{\nu\to 0}{\boldsymbol \tau}_w^\nu\not\equiv {\bf 0}.$

(i) If $\lim_{\nu\to 0}{\boldsymbol \tau}_w^\nu\equiv {\bf 0},$ then the infinite-Reynolds number limit coincides with the smooth Euler solution with the same initial data, as long as the latter exists. 
This is usually globally in time for potential flows.
In that case, there is at most a weak energy dissipation anomaly with exact potential-flow initial data,
although a strong dissipation anomaly might occur also if the initial potential flow is perturbed by 
turbulent, rotational fluctuations of small but non-vanishing energy. 

(ii) If $\lim_{\nu\to 0}{\boldsymbol \tau}_w^\nu\not\equiv {\bf 0},$ then the infinite-Reynolds number limit may be distinct from the 
smooth Euler solution with precisely the same initial data. A weak, singular Euler solution 
with a strong energy dissipation anomaly and vorticity cascade into the flow interior may 
coexist with the smooth Euler solution for the same initial data. 

However, the condition $\lim_{\nu\to 0}{\boldsymbol \tau}_w^\nu\not\equiv {\bf 0}$ requires very extreme near-wall events: non-zero
energy dissipation in a thin “Kato layer” (viscous sub-layer \& buffer layer) and 
discontinuity of the wall-normal velocity approaching the solid boundary through the 
inertial sublayer. The deterministic Navier-Stokes or any hydrodynamic description whatsoever
may break down, at least locally within the spacetime 
vicinity of the extreme event. 

Determining which of these possibilities is realized physically calls for a focused 
campaign of empirical investigation, both by numerical simulation and by laboratory experiment. 
Study of smoothly accelerated bodies looks especially promising, since this problem has obvious 
practical importance and yet exemplifies the fundamental issues. 
Computational efforts will need to take particular care for very fine space resolution near 
the body surface, for example with adaptive algorithms \cite{chatzimanolakis2022vortex}, 
in order to capture (or rule out) the emergence of the requisite small-scales. Since 
the events of interest may occur only sporadically in space-time, computational techniques 
for sampling extreme events \cite{farazmand2017variational,lestang2020numerical,rolland2022collapse}
may be useful. Experiments will be challenging also because of the sporadic nature of the events 
of interest and the lack of resolution of conventional measurement tools, such as particle-imaging 
velocimetry, near the body surface. New methods of measurement currently being developed \cite{orlu2020instantaneous,jassal2023accurate} might be crucial. Achieving high Reynolds 
numbers while maintaining hydraulic smoothness of the surface may require acceleration 
of large bodies. Laboratory experiments are, however, the only investigative tool that does 
not presuppose a particular mathematical model describing the flow and are thus indispensable.

\black{These issues have importance beyond the specific problem of external 
flow past bluff bodies. Onsager's analysis of the infinite Reynolds limit implies 
that observed anomalous dissipation of kinetic energy in high Reynolds turbulence 
requires suitable H\"older singularities of the limiting inviscid velocity field 
\cite{onsager1949statistical,eyink2024onsager} and experiments have provided 
evidence that the predicted singularities indeed exist 
\cite{lashermes2008comprehensive,debue2021three,faller2021nature}. However,
Onsager's theory leaves unexplained the origin of these singularities. 
A common view has been that singularities arise from finite-time blow-up of a smooth 
Euler solution (\cite{frisch1995turbulence}, \S 7.8). However, this mechanism 
is unavailable for 3D incompressible potential flows and for all 2D incompressible flows. 
Furthermore, such an origin of dissipative singularities would not account for the experimental 
observation that many flows exhibit anomalous dissipation only with hydraullically 
rough walls \cite{nikuradse1933laws,cadot1997energy}, in which pressure forces produce drag similarly 
as for bluff bodies. The conditional weak-strong uniqueness results for wall-bounded 
incompressible flows suggest a completely different origin of the required singularities: 
explosive events at the wall that lead in the inviscid limit to dissipative weak Euler solutions 
co-existing with the smooth Euler solution for the same initial-data. We remark again 
that the conditions that we established for weak-strong uniqueness in external flows 
past solid boidies are identical to those of Kelliher \cite{kelliher2017observations}
for certain classes of internal flows. In fact, experimental studies in the specific 
example of the von K\'arm\'an flow show that many of the points of higest local energy cascade 
and vorticity magnitude appear in close vicinity to the rotating impellers
that drive the flow (see \cite{faller2021nature}, Fig.5). It is reasonable to 
conjecture that the ultimate origin of dissipative singularities in 
high Reynolds-number incompressible fluid turbulence may lie in strong vorticity 
shed at solid walls during explosive separation.}

\section*{Acknowledgements} 
We thank T. Drivas for stressing the significance of ref.\cite{nguyenvanyen2018energy}. 
Our work was funded by the Simons Foundation, Targeted Grant MPS-663054 and Collaboration 
Grant MPS-1151713. G.E. thanks also 
the Department of Physics of the University of Rome `Tor Vergata' for hospitality 
while this work was finalized and acknowledges support from the European Research 
Council (ERC) under the European Union’s Horizon 2020 research and innovation program 
(Grant Agreement No. 882340).

\appendix 

\section{Navier-Slip Boundary Conditions}\lb{app:slip} 

It is well-known that stick boundary conditions at the wall, $\bu=\bzed$ on $\partial \Omega,$
although widely adopted, are only an approximation to more accurate Navier slip b.c. 
\be \bn\btimes(2\nu\bS\bdot\bn-\beta\bu)=\bzed, \quad \bu\bdot\bn=0, \lb{Nslip} \ee 
where $\beta$ is a friction coefficient (with units of velocity). 
In particular, such slip b.c. are necessary to model accurately a body impulsively accelerated 
through a molecular fluid, within a hydrodynamic framework \cite{reddy1967rayleigh}. However, 
the derivation of the 
Josephson-Anderson relation was previously explained using stick b.c. \cite{eyink2021josephson,eyink2021jaerratum}
and likewise its inviscid limit was demonstrated for those standard b.c. 
\cite{quan2024onsager}. To apply our weak-strong uniqueness result with the slip b.c.
\eqref{Nslip}, we must discuss briefly the slight changes required to those previous analyses.  

First, we recall the standard result from kinetic theory 
\cite{reichelsdorfer2016foundations,jiang2017boundary}
obtained already by Maxwell \cite{maxwell1879vii}, 
that slips arising from molecular effects have a {\it slip length} $b=\nu/\beta\sim \lambda_{mfp},$ 
the mean-free-path length, and friction coefficient $\beta\sim c_s,$ the sound speed. 
Because of the presence of the additional dimensional parameter $\beta$ (or $b$), usual Reynolds 
similarity breaks down. In fact, non-dimensionalizing the Navier-Stokes equations
with large scale length $L$ and velocity $U$ (outer units) yields slip b.c. 
\be \bn\btimes\left(\frac{2}{Re}\bS\bdot\bn-\frac{1}{Ma}\bu\right)=\bzed, \quad \bu\bdot\bn=0 \lb{nonNslip} \ee 
where in addition to Reynolds number $Re=UL/\nu$ there appears also the Mach number 
$Ma=U/c_s,$ which must be assumed sufficiently small for validity of the incompressible 
approximation. Since Reynolds similarity is broken, it now matters how the limit $Re\gg 1$ 
is achieved. In particular, increasing $U$ would eventually violate the condition
$Ma\ll 1,$ so that we consider instead decreasing $\nu$ or especially increasing $L$ 
with $U\ll c_s$ fixed. 

In mathematical parlance, we take dimensionless parameters $\nu=1/Re\to 0$
and $\beta=1/Ma$ fixed, which is known as the case of ``critical slip''. 
The theorems of \cite{drivas2018nguyen} still apply with Navier slip b.c., 
showing that inviscid limits yield dissipative weak Euler solutions under physically 
reasonable assumptions. Kato-type theorems have also been proved, at least  
for 2D domains \cite{wang2017vanishing}, 
implying strong $L^2$ convergence to the smooth Euler solution 
under conditions of vanishing dissipation in an $O(\nu)$-neighborhood of the boundary.  
Similar results have been proved without such conditions and assuming instead analytic 
initial data in a 2D flat wall geometry \cite{tao2020zero,nguyen2020inviscid}, but 
convergence holds only for a finite time $T>0$ over which analyticity is preserved. 
In fact, the case of a dipole vortex impinging on a 2D flat wall has been simulated 
numerically with critical Navier slip b.c. \cite{nguyen2011energy} and anomalous 
dissipation seems to appear after sufficiently long times, with apparent 
breakdown of weak-strong uniqueness for inviscid limits. 

The proof of the Josephson-Anderson relation 
\cite{eyink2021josephson,eyink2021jaerratum}, in fact,  does not 
differ for stick and slip b.c., because the only boundary condition used in 
the derivation are $\bu^\nu\bdot \bn=\bu_\phi\bdot \bn=0.$ The starting point is 
the force exerted on the body by the rotational flow, 
$$ {\bf F}_\omega^\nu=\rho\int_{\partial B} (-p_\omega^\nu\bn+\btau_w^\nu)\, dA$$
where 
\be \btau_w^\nu= 2\nu\bS^\nu\bdot\bn=\beta\bu^\nu\ee
is again a tangent vector field on the body surface. Standard arguments
\cite{lighthill1986informal,eyink2021josephson} yield the force-impulse relation 
${\bf F}_\omega^\nu=-d\bI_\omega^\nu/dt$ and then \cite{eyink2021jaerratum} the 
JA-relation: 
$${\mathcal W}_\omega^\nu(t)=- {\bf F}_\omega^\nu(t)\bdot\bV(t)= -\rho \int_\Omega
\bu_\phi\bdot(\bu^\nu\btimes\bomega^\nu-\nu\grad\btimes\bomega^\nu)\,dV. $$
Integration by parts with $\bu_\omega^\nu\bdot\bn=0$ yields
$${\mathcal W}_\omega^\nu(t)= -\rho \int_\Omega
\grad\bu_\phi\bdots\bu_\omega^\nu\bu_\omega^\nu\, dV +\eta\int_{\partial \Omega}
\bu_\phi\bdot(\bomega^\nu\btimes\bn)\,dA $$
but it is no longer true that $\nu\bomega^\nu\btimes\bn=\btau_w^\nu$ for slip b.c. 
In fact, a bit of computation shows that 
\be \nu\bomega^\nu\btimes\bn=\btau_w^\nu +2\nu (\grad\bn)\bdot\bu^\nu \lb{slip-omegn} \ee 
where $\bK=\grad\bn$ is the rank-2, symmetric $3\times 3$ matrix which defines the 
{\it Weingarten map} (or {\it shape map}) on the tangent space of the surface. 
Thus, for Navier slip b.c., there is an extra term in the JA-relation when 
written in this alternative form appropriate for taking the inviscid limit. Note however 
that the additional term vanishes as 
$\nu\to 0$ (or $Re\to\infty$) if $\bu^\nu\in L^2(0,T,L^2(\partial\Omega))$
uniformly in $\nu$. This can be expected from simple energy estimates, since $\beta$ is fixed 
as $\nu\to 0.$

Finally, we note that the global balance for the rotational flow energy is 
easily computed from Eq.(3.17) in \cite{eyink2021josephson} to be
\begin{eqnarray} 
&& \frac{dE_\omega^\nu}{dt} = -\eta\int_\Omega|\bomega^\nu|^2\, dV-\mu \int_{\partial\Omega}|\bu^\nu|^2\,dA \cr
&& -\int_{\Omega}{\boldsymbol\nabla}{\bf u}_{\phi}\,{\boldsymbol :} \, 
    {\bf u}^\nu_{\omega}\otimes{\bf u}^\nu_{\omega}\,dV +
    \int_{\partial\Omega} {\bf u}_\phi{\boldsymbol\cdot}{\boldsymbol\tau}_w^\nu \, dA \cr
&& \hspace{40pt} -2\eta\int_{\partial\Omega}\bu_\omega^\nu\bdot(\grad\bn)\bu^\nu\, dV \lb{slip-Eom} 
\end{eqnarray} 
where $\mu=\rho\beta$ and \eqref{slip-omegn} was used twice. The final term 
again is expected to vanish in the inviscid limit. The rigorous treaments 
of this limit in \cite{quan2024onsager,quan2025weak} carry through almost unchanged,
yielding a version of \eqref{rotbal} but with an additional dissipation term from surface 
friction. Thus, the condition $\bu_\phi\bdot\btau_w\equiv 0$
again suffices to derive weak-strong uniqueness (or, if necessary, one can also include the assumption 
that the final term in \eqref{slip-Eom} vanishes). However, since $\btau_w=\beta\bu$ with $\beta\neq 0$
in the limit $Re\to 0,$ it seems unlikely that $\bu_\phi\bdot\btau_w\equiv 0.$

\bibliography{bibliography.bib}
 
\end{document}